\documentclass[12pt]{article}
\usepackage[utf8]{inputenc}
\usepackage{lipsum}
\usepackage{hyperref}
\usepackage{enumitem}
\usepackage{float}
\usepackage{graphicx}
\usepackage{caption} 
\usepackage{calc}
\usepackage{fancyhdr}
\usepackage{tcolorbox}
\usepackage{hyperref}
\usepackage{wrapfig}
\usepackage{graphicx} 
\usepackage{amssymb}
\usepackage{amsmath}
\usepackage{multicol}
\usepackage{braket}
\usepackage{abstract}
\usepackage{multicol}
\usepackage{xcolor}
\title{\normalsize\bf Nonlinear Dynamics of the Inner Horizon in Reissner-Nordstr\"om\\ Black Holes: Insights into Mass Inflation}
\author{\normalsize {\sc Nihar Ranjan Ghosh\footnote{\tt g.nihar@iitg.ac.in}\ \ and Malay K. Nandy\footnote{\tt mknandy@iitg.ac.in}}\\
\normalsize \em Department of Physics, Indian Institute of Technology Guwahati\\
\normalsize \em Guwahati 781 039, India}
\date{\small (December 19, 2024)}
\begin{document}
\maketitle

\begin{abstract}
The well-known instability of the inner horizon of a Reissner-Nordstr\"om black hole, first suggested by Simpson and Penrose, although studied extensively, has remained illusive so far as several studies led to varied conclusions about the dynamical nature of the inner horizon. In this work, we therefore focus upon the dynamic nature of the inner horizon
in the course of mass inflation. We model this phenomenon with a massive chargeless scalar field minimally coupled with the Reissner-Nordstr\"om spacetime. Employing the Einstein-Maxwell field equation coupled with the Klein-Gordon equation, we obtain a nonlinear dynamical equation for the inner horizon coupled with the dynamics of the mass function and the scalar field. In the $S$-wave approximation, we develop a perturbative solution about the dynamic inner horizon and obtain an analytical solution as a polynomial of twelfth degree. Our detailed analysis shows that the inner horizon moves inward in the course of mass inflation. Higher the mass of the scalar field, faster are the shrinking rate of the inner horizon and the rate of mass inflation. Our solution for dynamic shrinking of the inner horizon suggests that a Reissner-Nordstr\"om spacetime tends towards a Schwarzschild-like geometry, in the infinite advanced time limit.
\end{abstract}


\section{\label{intro}Introduction}

Black holes, with their unique properties, are undoubtedly the most fascinating and remarkable outcomes of Einstein's general relativity. Understanding black hole physics is therefore of paramount importance, as they hold the key to unraveling the mysteries of nature. One particularly intriguing property is that certain types of black holes may enable multiverse travel. Potential candidates for such black holes include the Reissner–Nordstr\"om and Kerr black holes. However, these black holes also possess an instability inside their event horizon, known as \textit{mass inflation} \cite{poisson1989inner}, that hinders multiverse travel. Although numerous attempts have been made to determine the strength of the consequent singularity in various models, no definitive conclusion has been reached to date. To address this unresolved issue, this work focuses on a simple yet realistic model, designed to provide meaningful insights into the problem, specifically within the context of Reissner–Nordstr\"om geometry.

A Reissner-Nordstr\"om spacetime is formed when an electrically charged spherically symmetric mass distribution undergoes sufficient gravitational collapse. A static geometry of the spacetime is obtained as a unique solution of the Einstein-Maxwell equation \cite{reissner1916eigengravitation},\cite{nordstrom1918energy}. The spacetime metric for such geometry is given by
 \begin{equation}
\label{metric in t,r coordinate}
    ds^2=-f(r) dt^2+f^{-1}(r) dr^2+r^2 d\Omega^2
\end{equation}
where $f(r)=(1-\frac{2m_0}{r}+\frac{Q^2}{r^2})$ is the redshift function, with $m_0$ the ADM mass, $Q$ the electric charge of the black hole, and $d\Omega^2=d\theta^2+\sin^2\theta d\phi^2$. In this geometry, there are two horizons, given by $f(r)=0$,$$
r_{\pm}=m_0\pm\sqrt{m_0^2-Q^2}
$$
with $r_+$ and $r_-$ referred to as outer horizon and inner horizon, respectively. For $m_0=Q$, the two horizons coincide, resulting in an extremal black hole with only one horizon. For our work, we assume $m_0>Q$ initially, and study the behaviour of the inner horizon.

The inner horizon $r_-$ of the Reissner-Nordstr\"om black hole acts as a boundary of predictability because the zero of the redshift function $f(r)$ has a negative slope there. Therefore, any initial condition on any spacelike hypersurface will not be able to determine the geometry exactly beyond the inner horizon.
The redshift function is negative between $r_+$ and $r_-$, where the coordinates  $t$ and $r$ interchange their roles. 

The existence of a timelike singularity gives rise to a strange phenomenon, as was first discovered by Novikov. Novikov \cite{novikov1966change} first argued that any infalling passenger falling into the Reissner-Nordsrt\"om black hole, after crossing the inner horizon, can accelerate in the opposite direction to come out of the black hole into an asymptotically flat new universe without hitting the central singularity at $r=0$. 

However, the main issue with this multiverse travel is related to the fact that the inner horizon of the Reissner-Nordstr\"om black hole is unstable to any small external perturbation, as was first predicted by Penrose and Simpson \cite{simpson1973internal,Schpf1970BattelleR}. This prediction can be understood by considering any small external signal with equally spaced crests that will be highly blueshifted due to the strong gravitational field of the black hole resulting in an amplification of the signal energy. This amplification continues indefinitely at the inner horizon signifying its instability to any external perturbation \cite{hamilton2010physics}. 

In the Novikov scenario of the Reissner-Nordstr\"om black hole, the infalling passenger do not encounter any instability at the inner horizon and thus will be able to enter into an asymptotically flat universe. However, any physical black hole is prone to experiencing perturbations from external sources. Consequently, the entire concept of tunneling to another universe ultimately hinges on a single fundamental challenge: determining the nature and strength of the instability at the inner horizon under the most realistic situation.

In the case of rotating black holes, mass inflation poses similar issues, as can be seen from Refs. \cite{ori1992structure,bonanno1996mass,ori1997perturbative,hamilton2009interior}.

Several attempts have been made both in classical as well as semiclassical regime to address this problem. We describe below some of the attempts in the case of Reissner-Nordstr\"om black hole.

\subsection*{ Classical Mass Inflation}
 
Poisson and Israel  \cite{poisson1989inner} were the first to analyze the behavior of the mass function in Reissner-Nordstr\"om geometry, taking into account the back-reaction on the geometry caused by an infinitely blueshifted, spherically symmetric stream of infalling massless particles. Although this led to a charged Vaidya geometry, with an appropriate choice of coordinates, the instability was found to be weak. However, when the backscattered outgoing radiation was taken into account, the mass function's dependency on the advanced time $v$ changed significantly. They showed that the mass function $m(v)$ as measured by an infalling person will be of the form $m(U,V)\sim m_0\epsilon^2\delta(U)e^{k_0v}(v/m_0)^{-12}.$

Inclusion of the outgoing radiation causes this sudden and unbounded growth in the mass function. Because of this outgoing radiation, the inner apparent horizon separates from its Cauchy horizon. This separation occurs independently of the form of the outgoing radiation, as was first observed by Israel and Poisson \cite{poisson1990internal}.
 
In their model, the mass function $m(v)$ diverges exponentially, which they coined as \textit{mass inflation}. Additionally,  the Kretschmann scalar
 $R_{\mu\nu\alpha\beta}R^{\mu\nu\alpha\beta}\to\infty$ near the inner horizon, which is a stronger singularity compared to the charged Vaidya solution. Clearly, inclusion of the outgoing radiation modifies the strength of the singularity. The reason being the failure of the cross streams to move faster than light with respect to each other, which produces a large amount of gravitational energy that reveals itself as an infinite amount of local mass near the inner horizon \cite{brown2011mass}. 

Ori \cite{ori1991inner} improved the analysis by considering a shell of radiation, which divides the spacetime into two Vaidya geometry.  Matching the spacetime at the null shell gives the mass function as 
$m(v)\approx\Delta m(v)\propto|v|^{-1}(-\ln|v|)^{-p} $, where $p\ge12$ from Price's law\cite{price1972nonspherical}. Thus $m(v)$ and $R_{\mu\nu\alpha\beta}R^{\mu\nu\alpha\beta}\propto m^2$ both increase infinitely upon approaching the inner horizon. 

Furthermore, Ori showed that the growth rate of curvature and tidal forces are proportional to $\tau^{-2}|\ln|\tau||^{-p}$, where $\tau$ is the proper time. In addition, particle separations constituting an infalling object remain finite as $\tau\rightarrow0$, indicating a weak singularity. Therefore, even though there is a singularity at the inner horizon, its weakness suggests that the spacetime can be extended beyond inner horizon. This fact is further supported by the works of Tipler, \cite{tipler1977singularities} Ellis and Schmidt, \cite{ellis1977singular} and Burko \cite{burko1998analytic} on singularity strength criteria.

In our recent work \cite{ghosh2024mass}, we explored this issue in a more practical and realistic context. Specifically, we studied the dependency of the mass function $m(v)$ on advanced time $v$ of a Reissner-Nordstr\"om black hole subject to a perturbation through a massive scalar field. In doing so, we developed a perturbation series about the inner horizon, where the dynamics of the inner horizon was neglected for the sake of simplicity. This resulted in a double exponentially diverging mass function  of the form
$$m(v)\sim v^{-5}e^{-2\gamma v}\exp{[b\sigma v^2 e^{\gamma v}]}$$
where $b=(a^2+6a\beta+\beta^2)$ and $\gamma=a+\beta$, with $a=\mu^2 r_0^2$, $\sigma=\gamma/\beta$, $\beta=m_0/r_0^2$  are constants and $\mu$ is the mass of the scalar field. The double exponential behavior of the mass function indicates a much stronger singularity than those observed in earlier studies.

In fact, numerous studies have been conducted on this issue in the semiclassical framework as well, leading to different conclusions, as discussed below.

\subsection*{Semiclassical Mass Inflation}

Hiscock \cite{hiscock1977stress} demonstrated that the stress-energy tensor including quantum corrections, devoloped in the semiclassical framework by  Davies and Fulling \cite{davies1976energy},  diverges and hence  indicating an instability in the inner horizon of the Reissner-Nordstr\"om black hole.

In the semiclassical framework,  Balbinot and Poisson \cite{balbinot1993mass} studied the stress energy tensor and speculated that the mass inflation singularity may become stronger or regular depending on the sign of the quantum correction.   

On the other hand, Hwang and Yeom \cite{hwang2011internal} studied the interior of the Reissner-Nordstr\"om black hole during the formation and evaporation of the black hole in the semiclassical formalism. For small $Q/M$, they obtained a wormhole-like interior in the S-wave approximation, whereas mass inflation generates a large curvature in the interior during the final stages of evaporation.

In recent times, Barcel\'o  et al.~\cite{barcelo2021black, barcelo2022classical} studied the backreaction of a quantum field on the Reissner-Nordstr\"om geometry in the Polykov approximation. Incorporating the renormalized stress energy tensor of a massless quantized scalar field, they showed that the classical tendency of the inner horizon to move inward may be dominated by the semiclassical tendency to move outward. 

Carballo-Rubio et al.~\cite{carballo2022regular} and Franzin et al.~\cite{franzin2022stable} argued that regularized Reissner-Nordstr\"om black holes have vanishing surface gravity at the inner horizon. However, McMaken \cite{mcmaken2023semiclassical} proved that the instability at the inner horizon continues to exist.

\subsection*{de Sitter Models}

Brady and Poisson \cite{brady1992cauchy} perturbed the Reissner-Nordstr\"om black hole by an infalling stream of particles in the background of a de Sitter space representing the accelerated expansion of the Universe. They found mass inflation when the surface gravity at the inner horizon exceeds that of the cosmological horizon. However, for $|Q|>M$, the inner horizon remains stable for any generic perturbation \cite{brady1992cauchy, mellor1990stability}.

Moreover, Brady et al. \cite{brady1993cauchy} showed that the mass function remains bounded while the curvature diverges in a specific region of the parameter space. This prediction was confirmed by Cai and Su \cite{cai1995black} employing the stability conjecture of Helliwell and Konkowski \cite{helliwell1993testing, konkowski1994instabilities}. 
However, Markovi\'c and Poisson \cite{markovic1995classical} argued that the inner horizon must be unstable in the realm of quantum mechanics.

In the context of regularised black holes in classical gravity, the problem of mass inflation has been studied in Refs. \cite{bertipagani2021non,bonanno2023regular,carballo2021inner,bonanno2021regular}. See also a recent work in Ref. \cite{carballo2024mass}.

\subsection*{Present Work} 

The above discussion makes it obvious that various methodologies in classical and semiclassical frameworks have been employed to address the problem of mass inflation leading to varying conclusions. 
The common conclusion in all those works is that the inner horizon instability continues to exist and its strength appears to depend upon the model considered. As the strength of this singularity is the main concern, it is of utmost importance to check the behavior of the mass function in a more realistic scenario. 

All prior studies focused exclusively on massless scalar fields. However, since any physical black hole is likely to be surrounded by massive particles that continuously accrete into it, it is highly advisable to address this problem by considering perturbations caused by massive particles.

As discussed earlier, we previously addressed this problem by perturbing the black hole with a massive scalar field \cite{ghosh2024mass}. To the best of our knowledge, this was the first study involving a massive scalar field that successfully provided a closed analytical solution for the mass function. However, since the inner horizon radius must change as the mass of the black hole changes over time, a comprehensive analysis must also account for the dynamics of the inner horizon in order to gain a physically realistic and deeper understanding of the problem.

In this paper,  we therefore focus upon the dynamics of the inner horizon, in addition to  perturbing the Reissner-Nordstr\"om black hole with a massive, chargeless scalar field. Consequently, we develop a perturbation series about the dynamic inner horizon, and study the dynamical evolution of the inner horizon as well as the behaviour of the mass function with respect to the advanced time $v$. Due to the dynamical nature of the inner horizon, our findings show that the diverging nature of the mass function reduces significantly compared to the findings in the case of static horizon in \cite{ghosh2024mass}. 

The rest of the paper is organised as follows. In Section \ref{sec-2} we give a layout of our model of massive scalar field in the Reissner-Nordstr\"om background and present the Einstein-Maxwell field equations and the dynamics of the massive scalar field. Section \ref{sec-3} is the main focus of the paper where we develop a perturbation theory about the dynamical inner horizon in order to solve the coupled set of differential equations. Finally in Section \ref{sec-6}, we conclude the paper with a discussion of our results.

\section{The Model and Field Equations}
\label{sec-2}
As discussed above, we shall study the dynamical evolution of the inner horizon as well as the behaviour of the mass function with respect to the advanced time $v$ upon perturbing a Reissner-Nordstr\"om black hole by a massive, chargeless, scalar field $\Phi$. In doing so, we have to solve the coupled set of nonlinear dynamical equations originating from the appropriate action. We assume that at the initial advanced time $v_0$ the black hole has a unperturbed mass $m_0$ and corresponding inner horizon with radius $r_-=m_0-\sqrt{m_0^2-Q^2}$. 

We begin our analysis by modelling the situation with the combined action
\begin{equation}
    \label{action}
   S=S_R+S_\Phi,
\end{equation}
where
\begin{equation}
    S_R=\int d^4x \sqrt{-g}\left[\frac{1}{2}M_P^2 R-\frac{1}{4}F_{\mu\nu}F^{\mu\nu}    \right]
\end{equation}
and 
\begin{equation}
     S_\Phi
     =\int d^4x \sqrt{-g}\left[-\frac{1}{2}g^{\mu\nu}\nabla_\mu\Phi\nabla_\nu\Phi-\frac{1}{2}\mu^2\Phi^2\right]
\end{equation}
with $R$ the Ricci scalar, $\mu$ the mass of the scalar field $\Phi$, and $F_{\mu\nu}$ is the electromagnetic tensor. 

The Einstein field equation, as obtained from the above action (\ref{action}) is given by 
\begin{eqnarray}
    \begin{split}
         \label{Einstein}
     R_{\mu\nu}-&\frac{1}{2}g_{\mu\nu} R =\frac{1}{M_p^2}\left[F_{\nu\beta}F_\mu^\beta- \frac{1}{4}g_{\mu\nu}F_{\alpha\beta}F^{\alpha\beta}+\nabla_\mu\Phi\nabla_\nu\Phi-g_{\mu\nu}\left(\frac{1}{2}\nabla^\rho\Phi\nabla_\rho\Phi+\frac{1}{2}\mu^2\Phi \right)    \right] 
    \end{split}
\end{eqnarray} 

The spacetime being spherically symmetric, we represent its geometry by the metric $g_{\mu\nu}$ given by the line element  
\begin{equation}
    \label{metric in tortoise coordinate}
    ds^2=f(r)[-dt^2+dr^{*2}]+r^2 d\Omega^2,
\end{equation}
with the tortoise coordinate $r^*$ defined by $dr^*=f^{-1}dr$. 
For later convenience, we rewrite the line element as
\begin{equation}
\label{metric in v,r coordinate}
    ds^2=-f(r) dv^2+2dv dr+r^2 d\Omega^2,
\end{equation}
with $v$ the advanced time.

As a result of the perturbing scalar field, the black hole mass $m_0$ will be modified as $m_0\to m_0+m(v)$, which is a generalization of the Schwarzschild
mass and the Vaidya mass function to generic spherical
geometries, with the assumption of $m(v)=0$ without any perturbation when the spacetime reduces to the Reissner-Nordstr\"om black hole.  Therefore, 
\begin{equation}
f(v,r)=1-\frac{2\{m_0+m(v)\}}{r}+\frac{Q^2}{r^2}~.
\end{equation}

Equation (\ref{Einstein}) can be further simplified to obtain the dynamical equation for mass function $m(v)$ as 
\begin{equation}
    \begin{split}
        \label{Mass}
    \partial_am=4\pi r^2 T_a^b \partial_br
    \end{split}
\end{equation}
where $a,b=0,1$ and $T_{ab}$ being the energy momentum tensor corresponding to the chargeless, massive scalar field $\Phi$. This equations (\ref{Mass}) can be reexpressed as
\begin{equation}
    \begin{split}
        \label{mass}
    \frac{d m}{d v} =4\pi r^2& \left[   \left(\frac{\partial\Phi}{\partial v}  \right)^2+f(v,r)\left(\frac{\partial\Phi}{\partial v}  \right)\left(\frac{\partial\Phi}{\partial r}  \right)\right]~.
    \end{split}
\end{equation}

Similarly, the Klein-Gordon equation, $\left(\square-\mu^2\right)\Phi=0$, obtained from equation (\ref{action}), in the metric (\ref{metric in v,r coordinate}), can be expressed as
\begin{equation}
    \label{KG}
    \begin{split}
        2r^2\frac{\partial^2\Phi}{\partial v \partial r}+2r\frac{\partial\Phi}{\partial v}&+2rf\frac{\partial\Phi}{\partial r}+r^2\frac{\partial f}{\partial r}\frac{\partial\Phi}{\partial r}+r^2f\frac{\partial^2\Phi}{\partial r^2}-\mu^2r^2\Phi(v,r) =0,
    \end{split}
\end{equation}
describing the dynamics of the massive scalar field $\Phi$.

In order to study the behaviour of the mass function $m(v)$, we need to solve the two coupled, nonlinear, partial differential equations, given by equations (\ref{mass}) and (\ref{KG}).

\section{Perturbative Analysis at the Dynamical Horizon}
\label{sec-3}
    We make the assumption that the massive scalar field, $\Phi(r,v)$, is spherically symmetric (S-wave approximation) with its intensity falling off like $1/r^2$. Hence we have 
\begin{equation}
\label{Phi}
    \Phi(r,v)=\frac{1}{r}\phi(r,v)
\end{equation}
Substituting equation (\ref{Phi}) in equations (\ref{mass}) and (\ref{KG}) we get
\begin{equation}
\label{modified mass and KG equation}
    \begin{split}
        &\frac{dm}{dv} =4\pi\left[\left( \frac{\partial\phi}{\partial v} \right)^2+f\frac{\partial\phi}{\partial v}\left(\frac{\partial\phi}{\partial r}-\frac{\phi}{r}\right)\right]\\
        \end{split}
\end{equation}
and 
\begin{equation}
    \label{KG11}
    \begin{split}
         &rf\frac{\partial^2\phi}{\partial r^2}+2r\frac{\partial^2\phi}{\partial v\partial r}+r\frac{\partial f}{\partial r}\frac{\partial \phi}{\partial r}-\phi\frac{\partial f}{\partial r}-\mu^2r\phi =0\\
    \end{split}
\end{equation}

To explore the dynamics near the inner horizon, we expand  $f(r,v)$ and $\phi(r,v)$ about the dynamic inner horizon, defined by 
\begin{equation}
\label{rho}
\rho(v)=\big(m_0+m(v)\big)-\sqrt{\big(m_0+m(v)\big)^2-Q^2}, 
\end{equation}
as 
\begin{equation}
    \label{series expansion}
    \begin{split}
        f(r,v) &=f_0(v)+x f_1(v)+x^2 f_2
(v)+\dots\\
        \phi(r,v) &=\phi_0(v)+x\phi_1(v)+x^2\phi_2(v)+\dots\\
    \end{split}
\end{equation}
where  
\begin{equation}
\label{f0, f1, f2}
  \begin{split}
      &f_0(v)=0\\
      &f_1(v)=\Big(\frac{2\big(m_0+m(v)\big)}{\rho^2}-\frac{2Q^2}{\rho^3}\Big)\\
      &f_2(v)=\Big(-\frac{2\big(m_0+m(v)\big)}{\rho^3}+\frac{3Q^2}{\rho^4}  \Big)
  \end{split}  
\end{equation}
with $x=r-\rho(v)$.

Substituting equation (\ref{series expansion}) in equations (\ref{modified mass and KG equation}) and (\ref{KG11}) we get 
\begin{equation}
    \label{0th ordr mass eqn}
    \frac{1}{4\pi}\frac{dm}{dv}=\left(\frac{d\phi_0}{dv}-\frac{dr_0}{dv}\phi_1\right)^2
\end{equation}
and
\begin{equation}
    \label{0th ordr KG eqn}
    \begin{split}
        2\frac{d\phi_1}{dv}-4\phi_2 \frac{d\rho}{dv} +f_1\left(\phi_1-\frac{\phi_0}{\rho}\right)-\mu^2\phi_0=0
    \end{split}
\end{equation}
at order $\mathcal{O}(x^0)$, and
\begin{equation}
    \label{1st ordr mass eqn}
    \begin{split}
       2\frac{d\phi_1}{dv}-4\phi_2 \frac{d\rho}{dv} +f_1\left(\phi_1-\frac{\phi_0}{\rho}\right)=0
    \end{split}
\end{equation}
and
\begin{equation}
    \label{1st ordr KG eqn}
    \begin{split}
        4\frac{d\phi_2}{dv}+4\phi_2f_1+2f_2\phi_1-\frac{1}{\rho}&\left[f_1\left(\phi_1-\frac{\phi_0}{\rho}\right)+2\phi_0f_2\right]-\mu^2\phi_1=0\\
    \end{split}
\end{equation}
at order $\mathcal{O}(x^1)$.
Equations (\ref{0th ordr KG eqn}) and (\ref{1st ordr mass eqn}) indicate that $\phi_0(v)=0$.

\begin{figure}[H]
  \centering
    \includegraphics[scale=1.4]{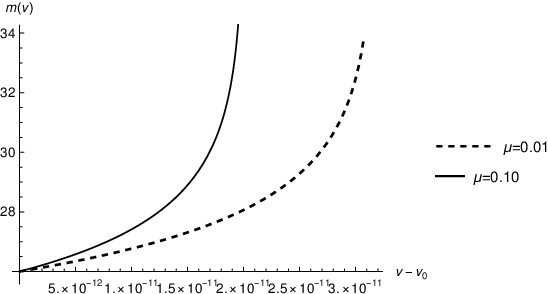}
    \caption{Evolution of the mass function $m(v)$ with respect to the advanced time $v$  as a consequence of the dynamical equation for $m(v)$ given by Equation (\ref{modified mass and KG equation}), obtained for scalar mass values $\mu=0.01,1.00$ in Planckian units.}   
    \label{figp2}
\end{figure}

Expanding the inner horizon $\rho(v)$ given by equation (\ref{rho}), we have
\begin{equation}
    \label{r0 taylr expnsn}
        \rho(v)=\frac{Q}{2}\left[\frac{Q}{\big( m_0+m(v)\big)}+\frac{Q^3}{4\big( m_0+m(v)\big)^3}+\frac{Q^5}{\big( m_0+m(v)\big)^5}+\dots \right].
\end{equation}
Assuming the black hole to be far from extreemality in the begining ($m_0\gg Q$), we can approximate $\rho(v)$ as
\begin{equation}
    \label{approximated r0(v)}
    \rho(v)\approx \frac{Q^2}{2\big( m_0+m(v)\big)}.
\end{equation}
This approximate expression for $\rho(v)$ becomes better and better with the progress of time $v$ as the mass function $m(v)$ is expected to increase due to mass inflation. We shall see that this expectation is consistent with our results of calculation, as shown later in Figure \ref{figp2}.

Using the  expression (\ref{approximated r0(v)}) for $\rho(v)$ in (\ref{f0, f1, f2}), we have 
\begin{equation}
    \label{apprx f1 f2}
    \begin{split}
        f_1(v)&=-\frac{Q^2}{\rho^3},\\
        f_2(v)&=\frac{2Q^2}{\rho^4}.\\
    \end{split}
\end{equation}

Thus from \ref{0th ordr mass eqn}, \ref{0th ordr KG eqn}, \ref{1st ordr mass eqn}, and \ref{1st ordr KG eqn}, we finally arrive at three coupled nonlinear differential equations for the dynamic inner horizon $\rho(v)$ and the scalar field $\phi$, given by 
\begin{equation}
    \label{r eqn}
    -\rho'(v)\rho^2(v)=\left(\frac{Q^2}{8\pi}\right)\frac{1}{\phi_1^2(v)},
\end{equation}
\begin{equation}
    \label{phi1 eqn}
    2\phi'_1(v)-4\rho'(v)\phi_2(v)+f_1(v)\phi_1(v)=0,
\end{equation}
and
\begin{equation}
    \label{phi2 eun}
    4\phi_2'(v)+4\phi_2(v)f_1(v)-\frac{5\phi_1(v)}{\rho(v)}f_1(v)-\mu^2\phi_1(v)=0,
\end{equation}
where a prime denotes derivative with respect to the advanced time coordinate $v$.

\begin{figure}[H]
  \centering
    \includegraphics[scale=1.4]{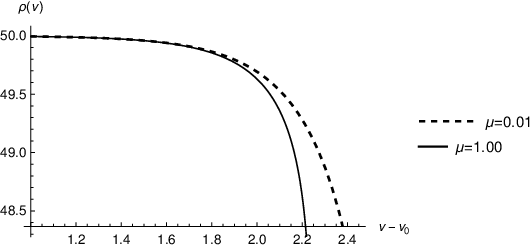}
    \caption{Evolution of the inner horizon $\rho(v)$ with respect to the advanced time $v$  as a consequence of the nonlinear dynamical equation for $\rho(v)$ given by Equation (\ref{r eqn}), obtained for scalar mass values $\mu=0.01,1.00$ in Planckian units.}   
    \label{figp1}
\end{figure}

Equation (\ref{phi2 eun}) can be rewritten in the form
\begin{equation}
    \label{phi2 apprx}
    \begin{split}
    4\frac{d}{dv}\left\{\phi_2\exp{\left(\int f_1 dv\right)}\right\}&=\exp{\left(-\int \frac{\mu^2}{5}\rho dv\right)}\frac{d}{dv}\left\{\frac{5\phi_1}{\rho}\exp{\int\left(\frac{\mu^2r_0}{5}+f_1 \right)dv}\right\}\\
    &-\exp{\left(-\int\frac{Q^2}{\rho^3}dv\right)}\frac{d}{dv}\left(\frac{5\phi_1}{\rho}\right)
     \end{split}
\end{equation}
Since the last term is an exponentially decreasing function of $v$, we can write 
\begin{equation}
    \begin{split}
        4\frac{d}{dv}&\left\{\phi_2\exp{\left(\int f_1 dv\right)}\right\}\approx \exp{\left(-\int \frac{\mu^2}{5}\rho dv\right)}\frac{d}{dv}\left\{\frac{5\phi_1}{\rho}\exp{\left[\int\left(\frac{\mu^2\rho}{5}+f_1 \right)dv\right]}\right\}\\  
        &=\exp{\left(-\int \frac{\mu^2}{5}\rho dv\right)} \frac{d}{dv}\left\{\frac{5\phi_1}{\rho}\exp{\left(\int f_1dv\right)}\right\}\exp{\left(\int\frac{\mu^2\rho}{5}dv\right)}\\
        &\hspace{0.5cm}+\left\{\frac{5\phi_1}{\rho}\exp{\left(\int f_1dv\right)}\right\}\frac{\mu^2\rho}{5}
    \end{split}
\end{equation}
in the late time approximation. We shall see that this approximation is consistent with our results of calculation, showing that $\rho\to 0$ as $v\to \infty$, as illustrated later in Figure \ref{figp1}.

Consequently,
\begin{equation}
\frac{d}{dv}\left[\left(4\phi_2-\frac{5\phi_1}{\rho}\right)\exp{\left(\int f_1dv\right)} \right]=\mu^2\phi_1\exp{\left(\int f_1dv\right)},
\end{equation}
giving
\begin{equation}
          4\phi_2=\frac{5}{\rho}\phi_1+\exp{\left(-\int f_1dv\right)}\int\mu^2\phi_1\exp{\left(\int f_1dv\right)}dv.
          \label{finl phi2}
\end{equation}

Substituting $\phi_2(v)$ from equation (\ref{finl phi2}) in equation (\ref{phi1 eqn}), we have
\begin{equation}
        2\phi_1'-\rho'\left[\frac{5}{\rho}\phi_1+\exp{\left(-\int f_1dv\right)}\int\left\{\mu^2\phi_1\exp{\left(\int f_1dv\right)}dv\right\}\right]+f_1\phi_1=0.
    \label{phi10}
\end{equation}
Multiplying throughout by $e^{\int f_1dv}$, and using the identity
\begin{equation}
        \phi_1'\exp{\left(\int f_1dv\right)}=\frac{d}{dv}\left\{\phi_1\exp{\left(\int f_1dv\right)}\right\}-f_1\phi_1\exp{\left(\int f_1dv\right)},
\end{equation}
we can write equation (\ref{phi10}) in the form
\begin{equation}
  \begin{split}
     2\frac{d}{dv}&\left\{\phi_1\exp{\left(\int f_1dv\right)}\right\}-2f_1\phi_1\exp{\left(\int f_1dv\right)} -5\frac{\rho'}{\rho}\phi_1\exp{\left(\int f_1dv\right)}\\
     &-\rho'\mu^2\int\left\{\phi_1\exp{\left(\int f_1dv\right)}\right\}dv+f_1\phi_1\exp{\left(\int f_1dv\right)}=0. 
  \end{split}
  \label{long eqn}
\end{equation}
We define $\phi_1e^{\int f_1dv}=\zeta(v)$, which simplifies equation (\ref{long eqn}) to 
\begin{equation}
    2\frac{d\zeta}{dv}-f_1\zeta-5\frac{\rho'}{\rho}\zeta-\rho'\mu^2\int\zeta dv=0.
\end{equation}

We further define $\int\zeta(v)dv=\eta(v)$, so that
\begin{equation}
    2\eta''-f_1\eta'-\frac{5\rho'}{r}\eta'-\rho'\mu^2\eta=0.
    \label{eta}
\end{equation}
To solve this differential equation, we substitute $\eta(v)=e^{\beta v}$, leading to
\begin{equation}
    \label{beta eqn}
     \exp{\left(\beta v-\int f_1dv\right)}=\rho^5\exp{\left(-\beta v-c_1+\frac{\mu^2\rho}{\beta}\right)},
\end{equation}
where $\beta c_1=-2\beta^2v_0+5\beta\ln{(r_-)}+\mu^2r_-$, with $c_1$ the integration constant, $r_-=m_0-(m_0^2-Q^2)^{1/2}$ being the radius of the inner horizon at the initial time $v=v_0$.

Now using the definations $\phi_1e^{\int f_1dv}=\zeta(v)$ and $\int\zeta(v)dv=\eta(v)=e^{\beta v}$, 
we obtain from equation (\ref{r eqn})
\begin{equation}
\label{rr eqn}
-\rho^2\rho'=\frac{c_2e^{\int2f_1dv}}{(\phi_1e^{\int f_1dv})^2}=\frac{c_2e^{\int2f_1dv}}{\zeta^2}=\frac{c_2e^{\int2f_1dv}}{\beta^2e^{2\beta v}}
\end{equation}
where $c_2=Q^2/(8\pi)$. Substituting equation (\ref{beta eqn}) in equation (\ref{rr eqn}), we have
\begin{equation}
    \label{r second eqn}
    -\int_{r_-}^\rho\rho^{12}\exp{\left[\frac{2\mu^2\rho}{\beta}\right]}d\rho=C\int_{v_0}^v \exp{[2\beta v]}dv
\end{equation}
with $C=c_2e^{2c_1}/\beta^2$.

Equation (\ref{r second eqn}) can be integrated to obtain the dynamics of the inner horizon $\rho(v)$ due to the massive chargeless scalar field. Subsequently, equation (\ref{approximated r0(v)}) can be used to obtain the dynamic behavior of the mass function $m(v)$.

Analytical solution of equation (\ref{r second eqn}) results in a $12$th order  polynomial in $\rho$ leading to the form 
\begin{equation}
    \label{final solution}
    \exp{\left[\frac{2\mu^2}{\beta}\rho(v)\right]}\sum_{n=0}^{12}a_n\rho^n(v)=\frac{C}{2\beta}\exp(2\beta v),
\end{equation}
where the coefficients $a_n$ result from integrating the left-hand side of equation (\ref{r second eqn}). 

The polynomial nature of the left-hand side of equation (\ref{final solution}) makes it impossible to invert this equation to get $\rho$ as a function of the advanced time coordinate $v$. Consequently, we obtain $\rho$ in terms of $v$ by numerical means.

Figure \ref{figp1} shows the time evolution of the inner horizon $\rho(v)$ as a function of $v$ for different masses of the scalar field, $\mu=0.10,1.00$. The figure clearly shows that the inner horizon shrinks more rapidly in the final stages as the mass of the scalar field increases. 

Figure \ref{figp2} shows the time evolution of the mass function $m(v)$ as a function of the advanced time $v$  for different masses of the scalar field, $\mu=0.10,1.00$. It is clearly seen that the mass function grows more rapidly in the final stages as the mass of the scalar field increases.

\section{Discussion and Conclusion}
\label{sec-6}

Ever since the instability of the inner horizon of a Reissner-Nordstr\"om black hole was first suggested by Simpson and Penrose \cite{simpson1973internal}, there have been a lot of effort in understanding the behavior of this instability, as discussed in Section \ref{intro}.   Considering the fact that these studies led to varied conclusions, it thus became necessary to probe the nature of the singularity. Moreover, there have been hardly any studies about the dynamic nature of the inner horizon, although the inner horizon must become dynamic in nature in the course of mass inflation.    

In this work, we therefore focused upon the dynamic nature of the inner horizon of a Reissner-Nordstr\"om black hole in the course of mass inflation. We modeled this situation with a massive chargeless scalar field minimally coupled with the Reissner-Nordstr\"om spacetime. As a result of the perturbation from the scalar field, the redshift function is modified to $f=1-2[m_0+m(v)]/r+Q^2/r^2$, giving a spherically symmetric metric with advanced time coordinate $v$. 

Employing the Einstein-Maxwell field equation coupled with the Klein-Gordon equation, we obtained the coupled dynamics of the mass function $m(v)$, the scalar field $\Phi(r,v)$, and the inner horizon $\rho(v)$ with respect to the advanced time $v$ [equations (\ref{mass}) and (\ref{KG})]. We took the $S$-wave approximation and the intensity of the scalar field was assumed to fall off like $1/r^2$.

In order obtain the dynamic nature of the inner horizon due to mass inflation, we developed a perturbative solution about the dynamic inner horizon in powers of $x=r-\rho(v)$, and obtained the coupled set of equations (\ref{0th ordr mass eqn},\ref{0th ordr KG eqn}) at order $\mathcal{O}(x^0)$ and equations (\ref{1st ordr mass eqn},\ref{1st ordr KG eqn}) at order $\mathcal{O}(x^1)$. Our detailed analysis shows that the scalar field $\phi(x,v)$ is obtained as a series expansion of the form 
\begin{equation*}
    \begin{split}
        \phi(x,v)=\phi_1&\Biggl[x+ x^2\left\{\frac{5}{4\rho(v)}+\frac{\mu^2}{4\phi_1\exp{\left(\int f_1dv\right)}}\int\phi_1\exp{\left(\int f_1dv\right)}dv \right\}\Biggl]+\mathcal{O}(x^3),
    \end{split}
\end{equation*}
in the asymptotic limit $v\to\infty$.

This ultimately yields a first-order differential equation, leading to the analytical  equation (\ref{r second eqn}) in integral form that captures the dynamical nature of the inner horizon $\rho(v)$ with respect to the advanced time $v$. Equation (\ref{r second eqn}) provides a polynomial expression for $\rho $ of order $12$, in the form
\begin{equation*}
    \exp{[\frac{2\mu^2}{\beta}\rho(v)]}\mathbb{P}_{12}(\rho)=\frac{C}{2\beta}\exp{[2\beta v]}
\end{equation*}
where $\mathbb{P}_{12}(\rho)$ is a $12$th order polynomial in $\rho$ whose exact form is found from equation (\ref{r second eqn}). 

The high order of the polynomial $\mathbb{P}_{12}(\rho)$ multiplied with the exponential function makes the inversion impossible to obtain the function $\rho=\mathcal{F}(v)$. This suggests that the only way to extract the behavior of $\rho(v)$ is through numerical analysis. 

As shown in Figure \ref{figp1}, numerical solution of equation (\ref{final solution}) shows that  the inner horizon of the Reissner-Nordstr\"om black hole moves inward, and $\rho\to 0^+$ in the asymptotic limit $v\to\infty$, due to mass inflation. In other words,  a Reissner-Nordstr\"om geometry tends towards a Schwarzschild geometry, due to the perturbation by a massive scalar field.  

As the inner horizon $\rho$ tends towards zero, equation (\ref{approximated r0(v)}) indicates 
\begin{equation*}
    m(v)=\left(\frac{Q^2}{2\rho(v)}-m_0\right)\to\infty
\end{equation*}
in the asymptotic limit $v\to\infty$.  Figure \ref{figp2} shows this behavior from the numerical solution.

Figure \ref{figp1} and \ref{figp2} further shows that the shrinking rate  of the inner horizon $\rho(v)$ towards zero, together with the rate of mass inflation, are more rapid as the mass of the scalar field  increases. 

These behaviors are quite anticipated since the scalar will be gravitationally blueshifted more strongly for higher masses than the lower ones.

\subsection*{Acknowledgement}
Nihar Ranjan Ghosh is supported by a Research  Fellowship from the Ministry of Human Resource Development (MHRD), Government of India.


\end{document}